\documentclass[review]{elsarticle}

\usepackage{lineno,hyperref}
\modulolinenumbers[5]

\journal{Journal of \LaTeX\ Templates}

\usepackage{bm}
\usepackage{mathrsfs}
\usepackage{amsmath}
\usepackage{amssymb}
\usepackage{graphicx}
\usepackage{array}
\usepackage{xcolor}
\usepackage{relsize}

\newcommand{\Rskin}[1]{$R_{\rm skin}^{#1}$}
\newcommand{\AlphaD}[1]{$\alpha_{\raisebox{1pt}{\tiny D}}^{#1}$}
\begin{document}
\begin{frontmatter}
\title{Electric dipole polarizability of neutron rich nuclei}
\author{J. Piekarewicz}
\address{Department of Physics \\ Florida State University \\
               Tallahassee, FL 32306, USA}
\ead{jpiekarewicz@fsu.edu}
\date{\today}
\begin{abstract}
 Insights into the equation of state of neutron rich matter obtained from the 
 neutron skin thickness of ${}^{208}$Pb are in sharp conflict with earlier 
 measurements of the electric dipole polarizability. We use a set of accurately 
 calibrated energy density functionals to highlight the tension and to articulate 
 how plans for a highly-intense Gamma Factory may alleviate the tension.
\end{abstract}
\end{frontmatter}

\smallskip


\section{Introduction}
\label{Sec:Introduction}

The quest to determine the equation of state (EOS) of neutron rich matter has been reenergized 
by recent astronomical discoveries that have opened the brand new era of multi-messenger 
astronomy\,\cite{Abbott:PRL2017,Cromartie:2019kug,Riley:2019yda,Miller:2019cac}. Also playing
a critical role in the determination of key parameter of the EOS are laboratory experiments at a 
variety of terrestrial facilities\,\cite{Thiel:2019tkm}. Experiments with electroweak probes provide 
the cleanest connection to the EOS and in particular to the slope of the symmetry energy at nuclear 
saturation density, a quantity often denoted by $L$. The symmetry energy quantifies the cost in turning 
symmetric nuclear matter into pure neutron matter. Experiments with electroweak probes constrain 
the symmetry energy by measuring the neutron skin thickness of 
${}^{208}$Pb (\Rskin{208})\,\cite{Abrahamyan:2012gp,Horowitz:2012tj,Adhikari:2021phr} and the 
electric dipole polarizability of a variety of nuclei\,\cite{Tamii:2013cna,Wieland:2009,Rossi:2013xha,
Hashimoto:2015ema,Birkhan:2016qkr}. Given the importance of the slope of the symmetry energy 
in the determination of several neutron-star properties---particularly stellar 
radii\,\cite{Lattimer:2006xb}---a concerted community effort has been devoted to a high-precision 
determination of $L$. Indeed, from a recent compilation of several theoretical predictions and 
experimental measurements the recommended value of $L$ is\,\cite{Drischler:2020hwi}:
\begin{equation}
 L\!=\! (59.8\pm4.1)\,{\rm MeV}.
 \label{Slope}
\end {equation} 
Such relatively small value indicates that the symmetry energy is soft, namely, that the pressure 
increases slowly with increasing density; see Fig.\,2 of Ref.\,\cite{Drischler:2020hwi}.

However, this finding has been recently brought into question by the PREX collaboration who has 
reported the following value for the neutron skin thickness of ${}^{208}$Pb\,\cite{Adhikari:2021phr}:
\begin{equation}
 \text{\Rskin{208}}=(0.283\pm0.071)\,{\rm fm}.
 \label{Rskin}
\end {equation} 
Using this value, plus invoking the strong correlation between \Rskin{208} and the following two 
key parameters of the symmetry energy, one obtains\,\cite{Reed:2021nqk}:
\begin{subequations}
\begin{align}
 & J = (38.1 \pm 4.7) {\rm MeV}, \\
 & L = (106 \pm 37) {\rm MeV},
\end{align} 
\label{JandL}
\end{subequations}
where $J$ is the symmetry energy at saturation density. The large discrepancy between this result for $L$
and the one quoted in Eq.(\ref{Slope}) highlights the tension that emerged after the PREX-2 measurement. 
Given that some of the earlier limits on $L$ were inferred from the analysis of the electric dipole polarizability 
of  ${}^{208}$Pb, we now proceed to confront those earlier results against the new limits inferred from 
invoking the newly recommended value for \Rskin{208}. From the experimental side, the commissioning 
of a Gamma Factory\,\cite{Placzek:2019xpw,Placzek:2020bjl} provides a unique opportunity to probe with 
unprecedented precision the electric dipole response of exotic, neutron rich nuclei.

\section{Formalism}
\label{Sec:Formalism}

The theoretical framework implemented in this work has been presented in much greater detail 
in several references, so we limit ourselves to highlight the most salient points; see for example
Ref.\,\cite{Piekarewicz:2013bea}. The starting point in our calculation of the nuclear response is 
a covariant energy density functional (EDF) accurately calibrated to properties of finite nuclei 
and neutron stars\,\cite{Chen:2014sca}. Once the calibration of the functional is completed, one 
proceeds to compute both ground-state properties as well as the linear response using a 
self-consistent formalism rooted in the random phase approximation (RPA). In the long wavelength 
limit, the distribution of isovector dipole strength $R(\omega;E1)$ is directly related to the 
photoabsorption cross section $\sigma_{\!\rm abs}(\omega)$. That is,
\begin{equation}
 \sigma_{\!\rm abs}(\omega) = \frac{16\pi^{3}}{9}\frac{e^{2}}{\hbar c}
 \omega R(\omega;E1).
\label{PhotoAbs}
\end{equation}
Previously identified as a strong isovector indicator\,\cite{Reinhard:2010wz}, the electric dipole 
polarizability is directly obtained from the inverse-energy-weighted sum of the dipole response:
\begin{equation}
\text{\AlphaD{}}  = \frac{\hbar c}{2\pi^{2}} \int_{0}^{\infty} 
\frac{\sigma_{\!\rm abs}(\omega)}{\omega^{2}}\,d\omega = 
\frac{8\pi e^2}{9} \int_{0}^{\infty}
\frac{R(\omega;E1)}{\omega}\,d\omega\;.
\label{alphaD}
\end{equation}
It is the powerful connection between the photoabsorption cross section, the electric dipole
polarizability, and the slope of the symmetry energy that make a highly intense Gamma Factory 
an ideal facility in the quest to constrain the equation of state.

\section{Results}
\label{Sec:Results}

In this section we present results for the electric dipole polarizability of ${}^{208}$Pb to 
highlight the disagreement of the PREX-2 informed prediction of \AlphaD{208} with 
an earlier measurement at the RCNP facility in Osaka, Japan\,\cite{Tamii:2013cna}. A 
high-precision measurement of the electric dipole response of ${}^{208}$Pb was obtained 
from a small-angle $({\bf p},{\bf p}')$ experiment that fully agreed with earlier photoabsorption 
measurements\,\cite{Tamii:2013cna}. The experimental distribution of strength contains a 
small non-resonant ``quasi-deuteron" (QD) contribution that must be removed before it can
be compared against theoretical (RPA) predictions. Once the QD contribution has been 
removed, one obtained the following estimate for the electric dipole polarizability of 
${}^{208}$Pb\,\cite{Roca-Maza:2015eza}:
\begin{equation}
  \text{\AlphaD{208}}\!=\!(19.6\pm0.6)\,{\rm fm}^{3}.
 \label{RCNPQD}
\end{equation}
\begin{figure}[ht]
 \centering
 \includegraphics[width=0.6\textwidth]{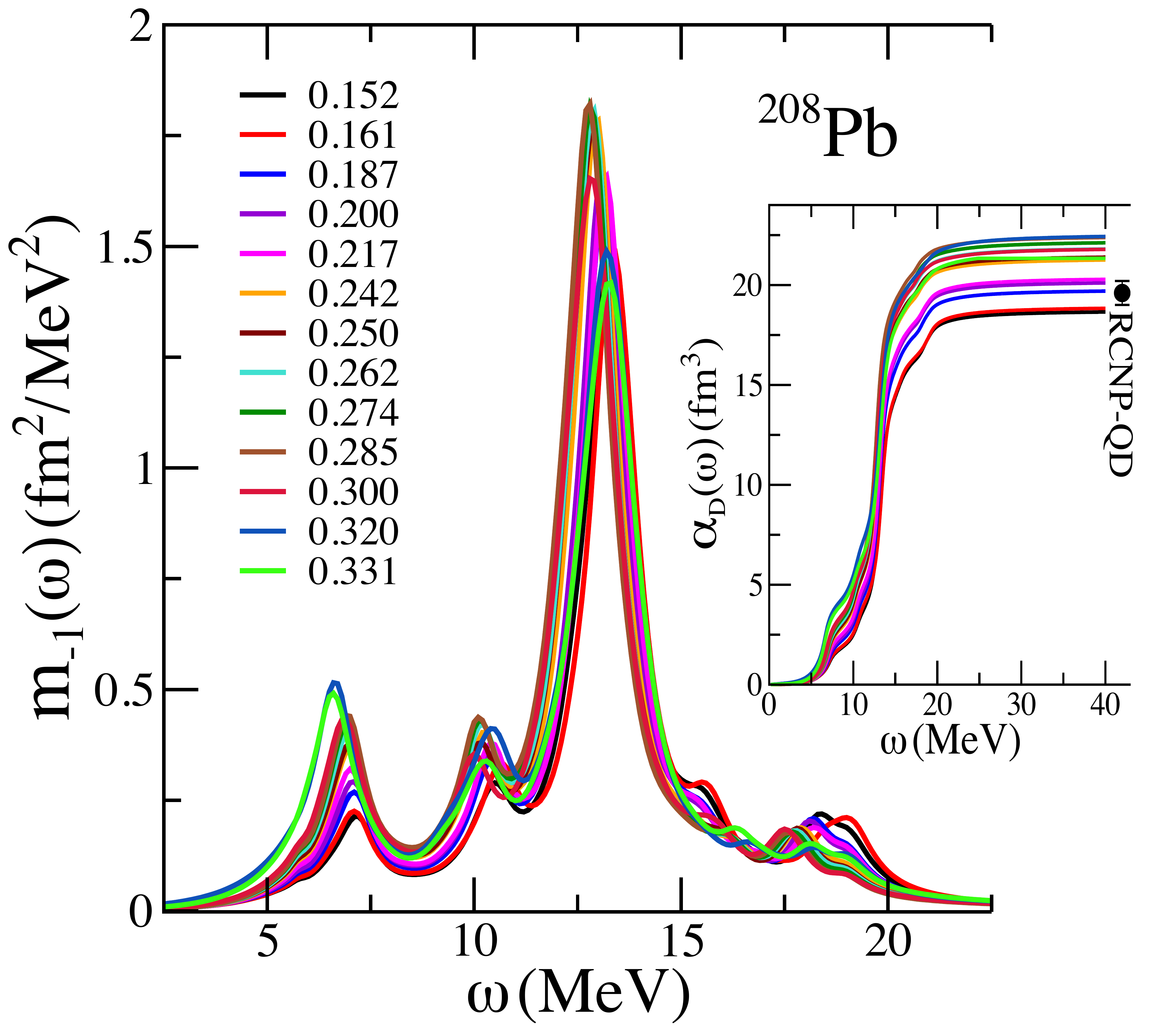}
 \caption{Distribution of electric dipole strength weighted by the inverse of the energy 
 for ${}^{208}$Pb as predicted by most of the covariant energy density functionals used 
 in Ref.\cite{Reed:2021nqk}. The labels denote the predicted value of \Rskin{208} for each 
 model and the inset displays the integrated strength, or ``running sum", with the value at 
 large excitation energy equal to \AlphaD{208}.}
\label{Fig1}
\end{figure}
To assess the impact of PREX-2 on \AlphaD{208} we display in Fig.\,\ref{Fig1} the 
inverse-energy-wighted distribution of electric dipole strength as predicted by a set of 13 covariant 
EDFs used in Ref.\,\cite{Reed:2021nqk}. The various functionals are consistent with ground state 
properties of final nuclei, yet are flexible enough in that they span a wide range of values for the 
neutron skin thickness of ${}^{208}$Pb, as indicated by the labels in the figure. This flexibility 
reflects our poor understanding of the density dependence of the symmetry energy. The distribution 
of strength displays a significant model dependence, that is better illustrated in the inset of 
Fig.\,\ref{Fig1}. The inset displays the ``running (or cumulative) sum" \AlphaD{}$(\omega)$, 
with the value at the largest excitation energy encoding the prediction for \AlphaD{208}. Note that
models with a stiffer symmetry energy, namely those that predict large values for both \Rskin{208}
and $L$, generate larger values for \AlphaD{208}\,\cite{Piekarewicz:2010fa}. Also included in the 
inset is the value extracted from the RCNP experiment, which favors smaller values for \AlphaD{208}
and therefore a relatively soft symmetry energy.

\begin{figure}[ht]
 \centering
 \includegraphics[width=0.6\textwidth]{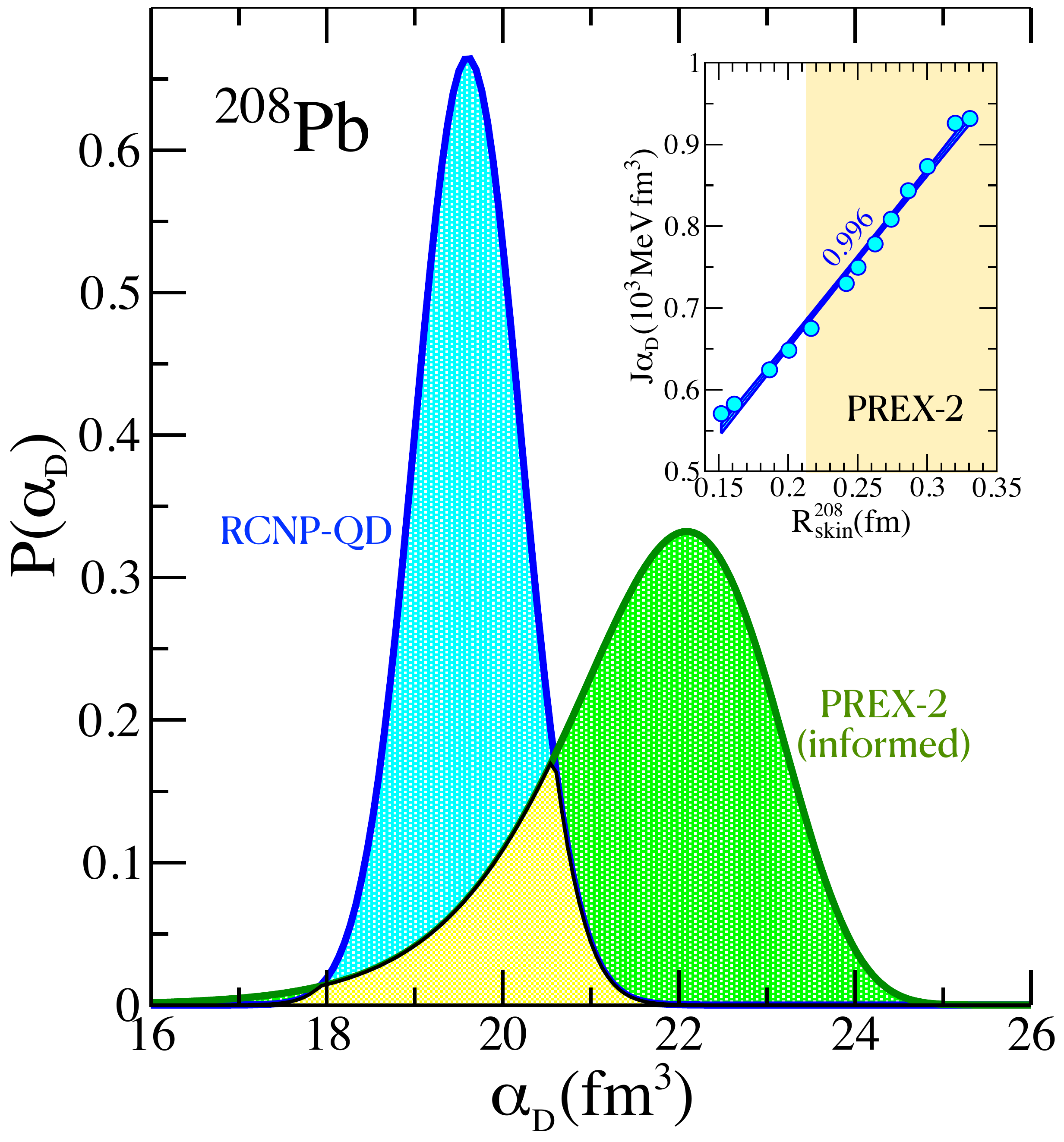}
 \caption{Probability distribution functions for the electric dipole polarizability of ${}^{208}$Pb.
 The normal distribution (RCNP-QD) was obtained from the experimental measurement of 
 \AlphaD{208}\,\cite{Tamii:2013cna} corrected by the removal of the quasi-deuteron  
 contribution\,\cite{Roca-Maza:2015eza}. The skew-normal distribution (PREX-2 informed) was 
 obtained from the constraints on both $J$ and $J$\AlphaD{208} imposed by the 
 PREX-2 result on \Rskin{208}\,\cite{Adhikari:2021phr}.}
\label{Fig2}
\end{figure}

Although the correlation between the electric dipole polarizability and the neutron skin thickness
is strong\,\cite{Reinhard:2010wz}, a far stronger correlation exists between \Rskin{208} and the 
product of $J$ times the electric dipole polarizability\,\cite{Satula:2005hy,Roca-Maza:2013mla}. Such 
a strong correlation is indicated in the inset of Fig.\,\ref{Fig2}, where the number above the line 
displays the correlation coefficient and the shaded region indicates the 1$\sigma$ error in 
\Rskin{208}\,\cite{Adhikari:2021phr}.  Also shown in Fig.\,\ref{Fig2} are the probability distribution 
functions for \AlphaD{208}\space obtained from both the RCNP experiment 
and the PREX-2-informed extraction. 
The skew-normal probability distribution informed by the PREX-2 result is obtained by relying on the 
nearly perfect correlation between $J$\AlphaD{208} and \Rskin{208} displayed in the inset---and also 
between $J$ and \Rskin{208} \,\cite{Reed:2021nqk}. In this manner one is able to generate two normal 
distributions, one for $J$ and another one for $J$\AlphaD{208}, from which the skew-normal distribution 
may be readily deduced. Following the standard practice of quoting a 68\% (1$\sigma$) estimate, we 
obtain the following value for the PREX-2-informed electric dipole polarizability:
\vspace{-5pt}
\begin{equation}
  \text{\AlphaD{208}}\!=\!(21.8^{+1.1}_{-1.4})\,{\rm fm}^{3},
 \label{AlphaDPrex2}
\end{equation}
where \AlphaD{208}$=\!21.8\,{\rm fm}^{3}$ is the median of the distribution. The figure clearly 
illustrates the incompatibility of the two results---at least at the 1$\sigma$ level. Indeed, the overlap 
region in Fig.\,\ref{Fig2}---estimated as the area shared by the two probability distributions---amounts 
to less than 25\%.  
\begin{figure}[ht]
 \centering
 \includegraphics[width=0.6\textwidth]{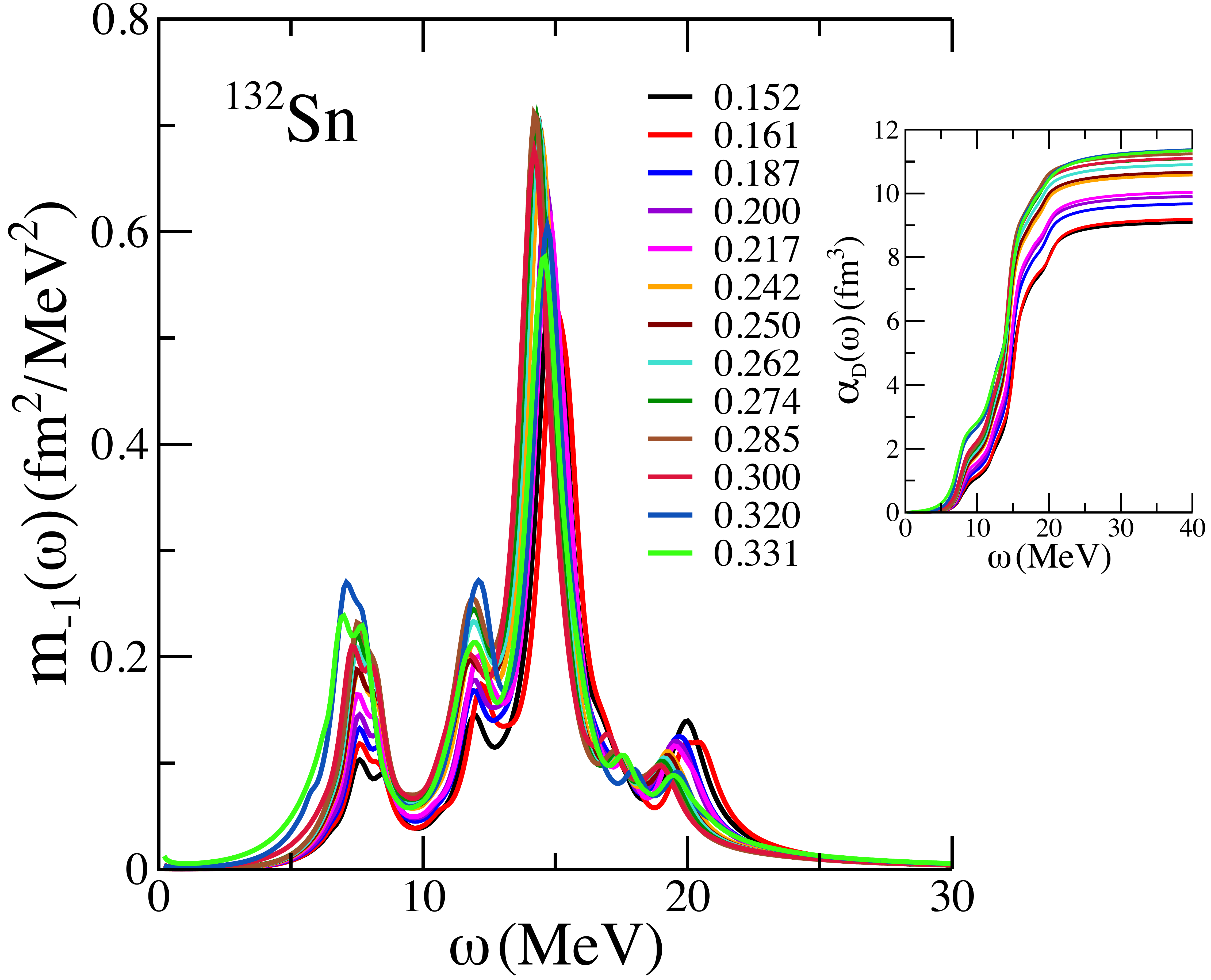}
 \caption{Distribution of electric dipole strength weighted by the inverse of the energy 
 for ${}^{132}$Sn as predicted by most of the covariant energy density functionals used 
 in Ref.\cite{Reed:2021nqk}. The labels denote the predicted value of \Rskin{208} for each 
 model and the inset displays the integrated strength, or ``running sum", with the value at 
 large excitation energy equal to \AlphaD{132}.}
\label{Fig3}
\end{figure}

We conclude the section by providing a prediction for the electric dipole polarizability of the unstable,
doubly-magic nucleus ${}^{132}$Sn. While a measurement of the electric dipole response of 
${}^{132}$Sn---both in the low- (``Pygmy") and high-energy (``Giant") regions---has been reported 
in Ref.\,\cite{Adrich:2005}, a precise value of the electric dipole polarizability is not yet available. 
Often portrayed as an oscillation of the excess neutrons against the isospin symmetric core, the 
emergence of a low-energy dipole mode has been found to correlate strongly with the development
of a neutron rich skin\,\cite{Piekarewicz:2006ip}. Indeed, the enhancement of the soft dipole 
mode is clearly discernible in Fig.\,\ref{Fig3}. This finding provides a compelling connection between 
the electric dipole polarizability and the neutron skin thickness---two critical observables used in the 
determination of the slope of the symmetry energy.

A precise measurement of the electric dipole polarizability in ${}^{132}$Sn is well motivated. It was observed
in Ref.\cite{Piekarewicz:2012pp} that the neutron skin thickness and the electric dipole polarizability
of ${}^{132}$Sn are both strongly correlated to the respective quantities in ${}^{208}$Pb. This suggests 
that a measurement of \Rskin{132} (if feasible) should reflect the PREX-2 result and yield a correspondingly
large neutron skin, suggesting that the symmetry energy is stiff. On the other hand, if one follows the 
\AlphaD{} correlation, then the relatively low value of \AlphaD{208} reported by the RCNP experiment 
suggests a correspondingly low value for ${}^{132}$Sn, and hence a soft symmetry energy.  The 
Gamma Factory could prove invaluable in solving this dilemma.

\section{Conclusions}
\label{Sec:Conclusions}

We have highlighted a tension between two different experimental techniques that are used to extract a
fundamental parameter of the equation of state: the slope of the symmetry energy $L$. Given that $L$ 
correlates strongly to a host of neutron-star observables, the resolution of the tension is of utmost importance. 
Indeed, laboratory experiments constitute the first rung in a ``density ladder" that aims to determine the
equation of state of neutron star matter. Naturally, one would like to see a significant reduction in the 
experimental uncertainty. However, the prospect of a more precise electroweak determination of \Rskin{208} 
in the near future is slim, given that the PREX campaign is now over. The prospects for improving the precision 
in the determination of the electric dipole polarizability are significantly better. The challenges in this arena 
are associated with the production of unstable nuclei with large neutron skins together with the determination 
of the electric dipole response over a wide range of energies. With photon fluxes that are more than 
a million times more intense than the HI$\gamma$S facility at the campus of Duke University, the Gamma 
Factory could play a vital role in the determination of the photoabsorption cross section of exotic nuclei with 
unprecedented precision. In so doing, the Gamma Factory could provide important constraints on the equation 
of state of neutron rich matter as we enter in earnest the era of muti-messenger astronomy.
 
\medskip
\section*{Acknowledgments}
 This material is based upon work supported by the U.S. Department of Energy Office of Science, 
 Office of Nuclear Physics under Award DE-FG02-92ER40750. 

\medskip
\section*{References}
\bibliography{GammaFactory.bbl}
\vfill\eject
\end{document}